\documentclass[preprint,showpacs,showkeys]{revtex4}
\usepackage{amssymb}

\begin{document}

\title{Legendre structure of the thermostatistics theory
based on the Sharma-Taneja-Mittal entropy}
\author{A.M. Scarfone}
\email{antonio.scarfone@polito.it} \affiliation{Istituto
Nazionale di Fisica della Materia (CNR-INFM) and
Dipartimento di Fisica,\\ Politecnico di Torino, Corso Duca
degli Abruzzi 24, 10129 Torino, Italy.}

\date{\today}

\begin {abstract}
The statistical proprieties of complex systems can differ
deeply for those of classical systems governed by
Boltzmann-Gibbs entropy. In particular, the probability
distribution function observed in several complex systems
shows a power law behavior in the tail which disagrees with
the standard exponential behavior showed by Gibbs
distribution. Recently, a two-parameter deformed family of
entropies, previously introduced by Sharma, Taneja and
Mittal (STM), has been reconsidered in the statistical
mechanics framework. Any entropy belonging to this family
admits a probability distribution function with an
asymptotic power law behavior. In the present work we
investigate the Legendre structure of the thermostatistics
theory based on this family of entropies. We introduce some
generalized thermodynamical potentials, study their
relationships with the entropy and discuss their main
proprieties. Specialization of the results to some
one-parameter entropies belonging to the STM family are
presented.
\end {abstract}

 \maketitle

%%%%%%%%%%%%%%%%%%%%%%%%%%%%%%%%%%%%%%%%%%%%%%%%%%%%%%%%%%%%%%%%%%%%%%%%%%%%%%%%%%%%%%

\section{Introduction}

Complex systems are ubiquitous in nature. Among the many,
we can quote high energy and nuclear physics, turbulence,
biophysics, geophysics, nano-systems, neural networks, but
also earthquakes and volcanic activity,  social sciences,
economic sciences and others
\cite{Abe0,Kaniadakis0,Kaniadakis00}. In general these
systems are governed by a nonlinear dynamics which
establishes a deep relation among the parts of the system
causing a strong correlation between them. As a consequence
these systems reach a dynamical equilibrium rather than a
statical equilibrium. Such equilibrium configuration (meta
equilibrium) changes slowly in time if compared to the time
scale of the underlying dynamic governing the system.
Remarkably, this dynamical equilibrium can deeply differ
from the statical equilibrium, in that the probability
distribution function (pdf) shows a behavior different from
the standard exponential one, typical of the Gibbs
distribution. In particular the statistical distribution
observed in several complex systems
exhibit a power law behavior in the tail.\\
A way to study the statistical proprieties of these {\em
anomalous} systems is based through the replacement of the
well-know Boltzmann-Gibbs (BG) entropy with a generalized
version of it \cite{Naudts, Scarfone1}. Very recently, a
generalized entropy with two parameters, previously
introduced in literature by Sharma and Taneja \cite{Sharma}
and Mittal \cite{Mittal}, in the framework of the
information theory, has been reconsidered in
\cite{Scarfone1} from a physical point of view. It has been
shown that the formulation of a statistical mechanics based
on the Sharma-Taneja-Mittal (STM) entropy preserve many
aspects of the theory based on the BG entropy. In
particular, it has been shown that it is positive definite,
continuous, symmetric, expandable, decisive, maximal,
concave, Lesche stable and fulfils a generalized Pesin-like
identity. In Ref. \cite{Scarfone3} the thermostatistics
proprieties of this
theory has been investigated in the microcanonical picture.\\
We remark that the STM entropy includes some one-parameter
entropies already investigated in literature, like the ones
introduced by Tsallis \cite{Tsallis}, by Abe \cite{Abe} and
by Kaniadakis \cite{Kaniadakis}. Consequently, it permits
us to consider all these
one-parameter entropies in a unified scheme.\\
The purpose of this work it to explore the Legendre
structure of the theory based on an entropy belonging to
the STM family. In the BG theory, through its Legendre
structure, we can introduce and establish some
relationships among different thermodynamical potentials
like, for instance, the free energy, the Massieus potential
and the entropy. It follows that the whole thermostatistics
theory can be formulated equivalently by using one or the
other of these quantities. In fact, depending by the
physical problem we have to deal with, each of these
potentials can be assumed as the more appropriate one for
the study of the
thermodynamical proprieties of the system \cite{Callen}.\\
It is then natural to ask if and how all these features can
be generalized in presence of an entropy belonging to the
STM family.

%%%%%%%%%%%%%%%%%%%%%%%%%%%%%%%%%%%%%%%%%%%%%%%%%%%%%%%%%%%%%%%%%%
\section{The Sharma-Taneja-Mittal entropy}

Let us recall the salient features of the STM entropy which
can be written in \cite{Scarfone1}
\begin{equation}
S_{_{\kappa,\,r}}=-\sum_ip_{_i}\,\ln_{_{\{\kappa,\,r\}}}(p_{_i}) \
,\label{stm}
\end{equation}
where $p\equiv\{p_{_i}\}_{_{i=1,\,\ldots,\,N}}$ is a
discrete pdf. Eq. (\ref{stm}) mimics the BG entropy through
the replacement of the standard logarithm with its
generalized version
\begin{equation}
\ln_{_{\{\kappa,\,r\}}}(x)=x^r\,\frac{x^\kappa-x^{-\kappa}}{2\,\kappa}
\ ,\label{log}
\end{equation}
where the deformation parameters $\kappa$ and $r$ are
restricted to the two dimensional region
$I\!\!R^2\supset{\mathcal R}=\{-|\kappa|\leq
r\leq|\kappa|,$ if $0\leq|\kappa|<1/2$ and
$|\kappa|-1\leq r\leq1-|\kappa|,$ if $1/2\leq|\kappa|<1\}$.\\
For any $(\kappa,\,r)\in{\mathcal R}$,
$\ln_{_{\{\kappa,\,r\}}}(x)=\ln_{_{\{-\kappa,\,r\}}}(x)$ is
a continuous, monotonic, increasing, concave and
normalizable function for $x\in I\!\!R^+$, with
$\ln_{_{\{\kappa,\,r\}}}(I\!\!R^+)\subseteq I\!\!R$. The
standard logarithm is recovered in the
$(\kappa,\,r)\to(0,\,0)$ limit:
$\ln_{_{\{0,\,0\}}}(x)\equiv\ln (x)$, as well as, in the
same limit, Eq. (\ref{stm}) converges to the BG entropy.
Moreover, Eq. (\ref{log}) satisfies the relation
$\ln_{_{\{\kappa,\,r\}}}(x)=-\ln_{_{\{\kappa,\,-r\}}}({1/
x})$ which, for $r=0$, reproduces the well known propriety
of the standard logarithm:
$\ln(x)=-\ln(1/x)$.\\
In Ref. \cite{Scarfone1}, starting from the following
functional-differential equation
\begin{equation}
\frac{d}{dx}\left[x\,\Lambda(x)\right]=\lambda\,\Lambda\left(\frac{x}{\alpha}\right)
\ ,\label{fde}
\end{equation}
obtained from certain physically justified assumptions, the
two-parameter deformed logarithm
$\Lambda(x)\equiv\ln_{_{\{\kappa,\,r\}}}(x)$ given in Eq.
(\ref{log}) has been derived, where the two constants
$\alpha$ and $\lambda$ are given by
\begin{equation}
\alpha=\left(\frac{1+r-\kappa}{1+r+\kappa}\right)^{1/2\,\kappa}
\
,\hspace{10mm}\lambda=\frac{(1+r-\kappa)^{(r+\kappa)/2\,\kappa}
}{(1+r+\kappa)^{(r-\kappa)/2\,\kappa}} \ ,\label{al}
\end{equation}
and are related in
$\ln_{_{\{\kappa,\,r\}}}\left(1/\alpha\right)=1/\lambda$.
The logarithmic solution fulfils the boundary conditions
$\Lambda(1)=0$ and $(d/dx)\,\Lambda(x)\Big|_{x=1}=1$.\\
Another solution of Eq. (\ref{fde}), $\Lambda(x)\equiv
u_{_{\{\kappa,\,r\}}}(x)$, with boundary conditions
$\Lambda(1)=1$ and $(d/dx)\,\Lambda(x)\Big|_{x=1}=r$, is
given by
\begin{equation}
u_{_{\{\kappa,\,r\}}}(x)=x^r\,\frac{x^\kappa+x^{-\kappa}}{2}
\ .\label{u}
\end{equation}
For any $(\kappa,\,r)\in{\mathcal R}$ the function
$u_{_{\{\kappa,\,r\}}}(x)=u_{_{\{-\kappa,\,r\}}}(x)$ is
continuous for $x\in I\!\!R^+$, with
$u_{_{\{\kappa,\,r\}}}(I\!\!R^+)\in I\!\!R^+$,
$u_{_{\{\kappa,\,r\}}}(0)=u_{_{\{\kappa,\,r\}}}(+\infty)=+\infty$
for $r\not=|\kappa|$ and it satisfies the relations
$u_{_{\{\kappa,\,r\}}}(x)=u_{_{\{\kappa,\,-r\}}}(1/x)$ and
$u_{_{\{\kappa,\,r\}}}\left(1/\alpha\right)=(1+r)/\lambda$.
Finally, it reduces to unity in the
$(\kappa,\,r)\to(0,\,0)$ limit:
$u_{_{\{0,\,0\}}}(x)=1$.\\Equations (\ref{log}) and
(\ref{u}) can be written in the form
\begin{equation}
\ln_{_{\{\kappa,\,r\}}}(x)={x^r\over\kappa}\,\sinh(\kappa\,\ln(x))
\
,\hspace{10mm}u_{_{\{\kappa,\,r\}}}(x)=x^r\,\cosh(\kappa\,\ln(x))
\ ,
\end{equation}
so that, many proprieties of $\ln_{_{\{\kappa,\,r\}}}(x)$
and $u_{_{\{\kappa,\,r\}}}(x)$ follow from the
corresponding ones of $\sinh(x)$ and $\cosh(x)$. For
instance, it is immediate to verify the following relations
\begin{eqnarray}
&&\ln_{_{\{\kappa,\,r\}}}(x\,y)=u_{_{\{\kappa,\,r\}}}(x)\,
\ln_{_{\{\kappa,\,r\}}}(y)+\ln_{_{\{\kappa,\,r\}}}(x)\,u_{_{\{\kappa,\,r\}}}(y)
\ ,\label{logxy}\\
&&u_{_{\{\kappa,\,r\}}}(x\,y)=u_{_{\{\kappa,\,r\}}}(x)\,u_{_{\{\kappa,\,r\}}}(y)
+\kappa^2\,\ln_{_{\{\kappa,\,r\}}}(x)\,\ln_{_{\{\kappa,\,r\}}}(y)
\ ,
\end{eqnarray}
as a consequence of the additivity formulae of the
hyperbolic functions.\\
In analogy with Eq. (\ref{stm}), for a given pdf, we
introduce the function
\begin{equation}
{\mathcal
I}_{_{\kappa,\,r}}=\sum_ip_{_i}\,u_{_{\{\kappa,\,r\}}}(p_{_i}) \
,\label{i}
\end{equation}
which can be seen as the linear mean value of the function
$u_{_{\{\kappa,\,r\}}}(x)$, according to the relation
${\mathcal I}=\langle u_{_{\{\kappa,\,r\}}}(p)\rangle$, as
well as the STM entropy can be defined as the linear mean
value of $-\ln_{_{\{\kappa,\,r\}}}(x)$, according to the
relation
$S_{_{\kappa,\,r}}=-\langle \ln_{_{\{\kappa,\,r\}}}(p)\rangle$.\\
We remark that Eq. (\ref{i}) reduces to unity in the
$(\kappa,\,r)\to(0,\,0)$ limit: ${\mathcal
I}_{_{0,\,0}}=\sum_ip_{_i}=1$ and likewise for an exact
distribution $p^{0}=\{0,\,\ldots,\,1,\,0,\,\ldots\}$:
${\mathcal
I}_{_{\kappa,\,r}}(p^{(0)})=1$.\\
The discrete pdf associated to the entropy (\ref{stm}),
under the constraints
\begin{equation}
\sum_ip_{_i}=1 \ ,\hspace{20mm}\sum_iE_{_i}\,p_{_i}=U \
,\label{const}
\end{equation}
on the normalization and on the linear mean energy $U$, can
be obtained through the following variational problem:
\begin{equation}
\frac{\delta}{\delta
p_{_j}}\left[-\sum_ip_{_i}\,\ln_{_{\{\kappa,\,r\}}}(p_{_i})
-\gamma\sum_ip_{_i}-\beta\sum_iE_{_i}\,p_{_i}\right]=0 \
,\label{var}
\end{equation}
where $\gamma$ and $\beta$, the Lagrange multipliers
associated to the constraints (\ref{const}), can be, in
case, obtained from Eqs. (\ref{const}) after we know the
$p_{_i}$.\\
Accounting for Eq. (\ref{fde}), from Eq. (\ref{var}) we
obtain
\begin{equation}
\lambda\,\ln_{_{\{\kappa,\,r\}}}\left(\frac{p_{_j}}{\alpha}\right)+\gamma+\beta\,E_{_j}=0
\ ,\label{var1}
\end{equation}
which gives the discrete pdf in the form
\begin{equation}
p_{_j}=\alpha\,\exp_{_{\{\kappa,\,r\}}}
\left(-\frac{\gamma+\beta\,E_{_j}}{\lambda}\right) \
.\label{pdf}
\end{equation}
In Eq. (\ref{pdf}) we have introduced the two-parameter
deformed exponential function
$\exp_{_{\{\kappa,\,r\}}}(x)$, the inverse function of
$\ln_{_{\{\kappa,\,r\}}}(x)$. Let us remark that, because
$\ln_{_{\{\kappa,\,r\}}}(x)$ is a strictly monotonic
function for any $(\kappa,\,r)\in{\mathcal R}$, its inverse
function certainly exists \cite{Scarfone1}.

%%%%%%%%%%%%%%%%%%%%%%%%%%%%%%%%%%%%%%%%%%%%%%%%%%%%%%%%%%%%%%%%%%%%%%
\section{Massieu functions and thermodynamical potentials}

In the classical thermostatistics, the thermodynamical
potentials are defined by means of Legendre transformation
on the mean energy. Although less known, another set of
functions can be introduced by performing a Legendre
transformation on the entropy. Such thermodynamical
potentials are named Massieu functions \cite{Callen}. For
instance, the free energy $F=U-S/\beta$ and the Massieu
function $\Phi=S-\beta\,U$ are obtained by means of
Legendre transformation on $U$ and $S$, respectively, and
are related each other through the relationship
$\Phi=-\beta\,F$. Let us explore such
Legendre structure for a theory based on the STM entropy.\\
We start by using in Eq. (\ref{var1}) the relation
(\ref{logxy}), with $x=p_{_i}$ and $y=1/\alpha$, so that it
follows
\begin{equation}
(1+r)\,\ln_{_{\{\kappa,\,r\}}}\left(p_{_j}\right)+u_{_{\{\kappa,\,r\}}}
\left(p_{_j}\right)+\gamma+\beta\,E_{_j}=0 \ .\label{var2}
\end{equation}
By taking the average of Eq. (\ref{var2}) with respect to
$p_{_i}$ we obtain
\begin{equation}
S_{_{\kappa,\,r}}={1\over1+r}\,\left({\mathcal
I}_{_{\kappa,\,r}}+\gamma+\beta\,U\right) \ ,\label{s1}
\end{equation}
which recovers, in the $(\kappa,\,r)\to(0,\,0)$ limit, the
classical
relationship $S=1+\gamma+\beta\,U$.\\
From Eq. (\ref{s1}) we can derive some useful proprieties
concerning the STM entropy.\\
For instance, by recalling Eq. (\ref{fde}), we can write
\begin{equation}
\frac{d\,S_{_{\kappa,\,r}}}{d\,U}=-\sum_i\frac{d}{dp_{_i}}\left[p_{_i}\,\ln_{_{
\{\kappa,\,r\}}}\left(p_{_i}\right)\right]\,\frac{dp_{_i}}{dU}
=-\lambda\sum_i \ln_{_{
\{\kappa,\,r\}}}\left(\frac{p_{_i}}{\alpha}\right)\,\frac{dp_{_i}}{dU}
\ ,\label{ds}
\end{equation}
and taking into account the expression of the $p_{_i}$ we
obtain
\begin{equation}
\frac{d\,S_{_{\kappa,\,r}}}{d\,U}=\sum_i
\left(\gamma+\beta\,E_{_i}\right)\,\frac{dp_{_i}}{dU} \
.\label{ds1}
\end{equation}
By assuming the ``no work'' condition
$\sum_ip_{_i}\,dE_{_i}=0$, which implies
$dU=\sum_iE_{_i}\,dp_{_i}$, and taking into account that
$\sum_idp_{_i}=0$, as it follows from the normalization on
$p_{_i}$, we obtain
\begin{equation}
\frac{d\,S_{_{\kappa,\,r}}}{d\,U}=\beta \ .\label{ds2}
\end{equation}
From this equation we see that $S_{_{\kappa,\,r}}$ is a
function of the mean energy $U$ and that, like in the BG
theory, $\beta$ and $U$ are
variables canonically conjugated.\\
The generalized Massieu potential $\Phi_{_{\kappa,\,r}}$
can be introduced by performing a Legendre transformation
on the entropy:
\begin{equation}
\Phi_{_{\kappa,\,r}} =
S_{_{\kappa,\,r}}-\frac{d\,S_{_{\kappa,\,r}}}{d\,U}\,U
\equiv S_{_{\kappa,\,r}}-\beta\,U \ ,\label{fl}
\end{equation}
and after using Eq. (\ref{s1}) in Eq. (\ref{fl}) we obtain
\begin{equation}
\Phi_{_{\kappa,\,r}}={1\over1+r}\left({\mathcal
I}_{_{\kappa,\,r}}+\gamma-r\,\beta\,U\right) \ .\label{f}
\end{equation}
It is trivial to verify the validity of the relation
\begin{equation}
\frac{d\,\Phi_{_{\kappa,\,r}}}{d\,\beta}=-U \ ,\label{ds3}
\end{equation}
as it follows readily by using Eqs. (\ref{ds2}) and
(\ref{fl}). Equation (\ref{ds3}) still states that $\beta$
and $U$ are canonically conjugated variables and that
$\Phi_{_{\kappa,\,r}}$ is, as matter of fact, a function of $\beta$. \\
Finally, we observe that, if one is welling to considers
the free energy as a function of $1/\beta$, the generalized
free energy $F_{_{\kappa,\,r}}$ can be introduced, through
a Legendre transformation on $U$:
\begin{equation}
F_{_{\kappa,\,r}}=
U-\frac{d\,U}{d\,S_{_{\kappa,\,r}}}\,S_{_{\kappa,\,r}}
\equiv U-{1\over\beta}\,S_{_{\kappa,\,r}} \ .\label{f1}
\end{equation}
By using Eq. (\ref{s1}) in Eq. (\ref{f1}) we obtain
\begin{equation}
F_{_{\kappa,\,r}}=-{{\mathcal
I}_{_{\kappa,\,r}}+\gamma-r\,\beta\,U\over(1+r)\,\beta} \
.\label{f}
\end{equation}
Moreover, it results
\begin{equation}
F_{_{\kappa,\,r}}=-{\Phi_{_{\kappa,\,r}}\over\beta} \ ,
\end{equation}
like in the standard thermostatistics theory, from which it
is easy to show that
\begin{equation}
\frac{d\,F_{_{\kappa,\,r}}}{d(1/\beta)}=-S_{_{\kappa,\,r}}
\ ,
\end{equation}
imitating in this way the classical relationships between
the free energy and the entropy.

%%%%%%%%%%%%%%%%%%%%%%%%%%%%%%%%%%%%%%%%%%%%%%%%%%%%%%%%%%%%%%%%%%%%%%
\section{Canonical partition function}
Let us introduce the generalized canonical partition
function $Z_{_{\kappa,\,r}}$ through the relation
\begin{equation}
\ln_{_{\{\kappa,\,r\}}}\Big(Z_{_{\kappa,\,r}}\Big)={1\over
1+r}\,\left({\mathcal
I}_{_{\kappa,\,r}}+\gamma-r\,\beta\,U\right) \ ,\label{z}
\end{equation}
so that, by inserting Eq. (\ref{z}) into Eq. (\ref{s1}) we
obtain
\begin{equation}
S_{_{\kappa,\,r}}=\ln_{_{\{\kappa,\,r\}}}\Big(
Z_{_{\kappa,\,r}}\Big)+\beta\,U \ ,\label{ddz}
\end{equation}
which mimics the standard relation $S=\ln(Z)+\beta\,U$.\\
We notice that some authors prefer to introduce a partition
function $\overline Z_{_{\kappa,\,r}}$ which refers to the
energy levels $\{E_{_i}\}$ with regards to $U$
\cite{Tsallis1}. The two different definitions are related
each to the other by
\begin{equation}
\ln_{_{\{\kappa,\,r\}}}\Big(\overline
Z_{_{\kappa,\,r}}\Big)=\ln_{_{\{\kappa,\,r\}}}\Big(
Z_{_{\kappa,\,r}}\Big)+\beta\,U \ ,\label{mz}
\end{equation}
so that Eq. (\ref{ddz}) simplifies in
$S_{_{\kappa,\,r}}=\ln_{_{\{\kappa,\,r\}}}\left(\overline
Z_{_{\kappa,\,r}}\right)$.\\
 In order to verify the consistence of the definition
(\ref{z}) let us evaluate the following derivative
\begin{eqnarray}
\frac{d\,{\mathcal I}_{_{\kappa,\,r}}}{d\,\beta}=
\sum_i\frac{d}{d\,p_{_i}}\left[p_{_i}\,u_{_{\{\kappa,\,r\}}}
\left(p_{_i}\right)\right]\,\frac{d\,p_{_i}}{d\,\beta} =
\lambda\sum_iu_{_{\{\kappa,\,r\}}}
\left(\frac{p(x_{_i})}{\alpha}\right)\,\frac{d\,p(x_{_i})}{d\,\beta}
\ ,\label{di}
\end{eqnarray}
where we have posed $p(x_{_i})\equiv p_{_i}$ and
$x_{_i}=\gamma+\beta\,E_{_i}$. By using the relation
\begin{equation}
\lambda\,u_{_{\{\kappa,\,r\}}}
\left(\frac{p(x_{_i})}{\alpha}\right)\,d\,p(x_{_i})=r\,x_{_i}\,dp(x_{_i})
-p(x_{_i})\,dx_{_i} \ ,
\end{equation}
as it follows by deriving Eq. (\ref{pdf}), we can rewrite
\begin{eqnarray}
\nonumber \frac{d\,{\mathcal
I}_{_{\kappa,\,r}}}{d\,\beta}&=&
r\sum_ix_{_i}\,\frac{d\,p(x_{_i})}{d\,\beta}-\sum_i
p(x_{_i})\,\frac{d\,x_{_i}}{d\,\beta}\\
\nonumber&=&
r\,\frac{d}{d\,\beta}\left(\sum_ip(x_{_i})\,x_{_i}\right)-(1+r)\sum_ip(x_{_i})\,
\frac{d\,x_{_i}}{d\,\beta}\\
\nonumber
&=&r\,\frac{d}{d\,\beta}\left(\gamma+\beta\,U\right)-(1+r)
\frac{d\,\gamma}{d\,\beta}-(1+r)\,U\\
&=&r\,\beta\,\frac{d\,U}{d\,\beta}-\frac{d\,\gamma}{d\,\beta}-U
\ .\label{di1}
\end{eqnarray}
On the other hand, by taking the derivative of Eq.
(\ref{z}) with respect to $\beta$, we obtain
\begin{equation}
\frac{d}{d\,\beta}\,\ln_{_{\{\kappa,\,r\}}}\Big(
Z_{_{\kappa,\,r}}\Big)={1\over
1+r}\,\left(\frac{d\,{\mathcal
I}_{_{\kappa,\,r}}}{d\,\beta}+\frac{d\,\gamma}{d\,\beta}
-r\,U-r\,\beta\,\frac{d\,U}{d\,\beta}\right) \ ,\label{di2}
\end{equation}
and accounting for Eq. (\ref{di1}) it follows
\begin{equation}
\frac{d}{d\,\beta}\,\ln_{_{\{\kappa,\,r\}}}\Big({
Z_{_{\kappa,\,r}}}\Big)=-U \ ,\label{dz}
\end{equation}
according to the classical relationship $d\,\ln(Z)/d\,\beta=-U$.\\
We remark that, by comparing Eq. (20) with Eq. (\ref{z}) it
follows
\begin{equation}
\Phi_{_{\kappa,\,r}}=\ln_{_{\{\kappa,\,r\}}}\Big(
Z_{_{\kappa,\,r}}\Big)
\end{equation}
so that, accounting for Eq. (\ref{dz}), we recover again Eq. (\ref{ds3}).\\
Finally, we recall that in the classical statistical
mechanics the canonical partition function encodes all the
statistical proprieties of the system. This feature also
holds in the generalized theory under investigation. In
fact, by assuming $\beta\simeq constant$ for a long period
of time, we have
\begin{equation}
\frac{d}{d\,E_{_i}}\,\ln_{_{\{\kappa,\,r\}}}\Big(
Z_{_{\kappa,\,r}}\Big)={1\over
1+r}\,\left(\frac{d\,{\mathcal
I}_{_{\kappa,\,r}}}{d\,E_{_i}}+\frac{d\,\gamma}{d\,E_{_i}}
-r\,\beta\,\frac{d\,U}{d\,E_{_i}}\right) \ ,\label{die}
\end{equation}
and following the same argument used in Eq. (\ref{di1}) it
follows
\begin{equation}
\frac{d\,{\mathcal I}_{_{\kappa,\,r}}}{d\,E_{_i}}=-(
1+r)\,\beta\,p_{_i}-\frac{d\,\gamma}{d\,E_{_i}}
+r\,\beta\,\frac{d\,U}{d\,E_{_i}} \ ,
\end{equation}
so that, from Eq. (\ref{die}), we obtain
\begin{equation}
p_{_i}=-{1\over\beta}\,\frac{d}{d\,E_{_i}}
\,\ln_{_{\kappa,\,r}}\Big(Z_{_{\kappa,\,r}}\Big) \ ,
\end{equation}
i.e., the equilibrium distribution can be derived
equivalently through the generalized canonical partition
function.

%%%%%%%%%%%%%%%%%%%%%%%%%%%%%%%%%%%%%%%%%%%%%%%%%%%%%%%%%%%%%%%%%%%%%%
\section{Particular cases}

Let us specify our results to some relevant one-parameter
deformed entropies belonging to the STM family.\\
As a first example, we choose $r=0$. From Eq. (\ref{stm})
we obtain the entropy proposed by Kaniadakis
\cite{Kaniadakis}:
\begin{equation}
S_{_\kappa}=-\sum_i\frac{p_{_i}^\kappa-p_{_i}^{-\kappa}}{2\,\kappa}
\ .\label{sk}
\end{equation}
It was conjectured that this entropy emerges naturally in
the context of the special relativity. Among the many
possible applications, the entropy (\ref{sk}) has been
employed in the reproduction of the energy distribution of
the fluxes of cosmic rays \cite{Kaniadakis} and in the
study of the fracture propagation in brittle materials
\cite{Scarfone5}, showing a good agreement with the data
observed both experimentally and through
numerical simulation.\\
Starting from Eq. (\ref{sk}) and from its related function
\begin{equation}
{\cal
I}_{_\kappa}=\sum_i\frac{p_{_i}^\kappa+p_{_i}^{-\kappa}}{2}
\ ,
\end{equation}
we can define the canonical $\kappa$-partition function
$Z_{_\kappa}$ as
\begin{equation}
\ln_{_{\{\kappa\}}}\Big(Z_{_\kappa}\Big)=
S_{_\kappa}-\beta\,U\equiv{\cal I}_{_\kappa}+\gamma \ .
\end{equation}
Remark that, by introducing the function $\overline
Z_{_\kappa}$, by means of Eq. (\ref{mz}), we obtain the
relation
\begin{equation}
\left(\overline
Z_{_\kappa}\right)^\kappa=\kappa\,S_{_\kappa}+\sqrt{1+\kappa^2\,S_{_\kappa}^2}
\ .
\end{equation}
Finally, the expressions of the $\kappa$-Massieu potential
$\Phi_{_\kappa}$ and the $\kappa$-free energy $F_{_\kappa}$
are given, respectively, by
\begin{eqnarray}
&&\Phi_{_\kappa}=
\ln_{_{\{\kappa\}}}\Big(Z_{_\kappa}\Big)\equiv{\cal
I}_{_\kappa}+\gamma \ ,\\
&&F_{_\kappa}=-{1\over\beta}\,\ln_{_{\{\kappa\}}}\Big(Z_{_\kappa}\Big)\equiv-{1\over\beta}\Big({\cal
I}_{_\kappa}+\gamma\Big) \ .
\end{eqnarray}

As a second example we pose in Eq. (\ref{stm})
$r=\pm|\kappa|$ and after introducing the parameter
$q=1\mp2\,|\kappa|$ we obtain
\begin{equation}
S_{_{2-q}}=\sum_i\frac{p_{_i}^{2-q}-p_{_i}}{q-1} \
,\label{Tsallis}
\end{equation}
which coincides with the Tsallis entropy in the ``$2-q$
formalism'' \cite{Scarfone4}.\\ After its introduction,
entropy (\ref{Tsallis}) has been widely applied, as a
paradigm, in the study of the statistical proprieties of
complex systems shoving a pdf
with a power law behavior in the tail \cite{Tsallis}.\\
The canonical $q$-partition function $Z_{_{2-q}}$
associated to the entropy (\ref{Tsallis}) is defined by
\begin{equation}
\ln_{_{2-q}}\Big(Z_{_{2-q}}\Big)=
S_{_{2-q}}-\beta\,U\equiv{2\over3-q}\Big({\cal
I}_{_{2-q}}+\gamma\Big)+{q-1\over3-q}\,\beta\,U \
,\label{zq}
\end{equation}
where the function ${\cal I}_{_{2-q}}$ takes the expression
\begin{equation}
{\cal I}_{_{2-q}}=\sum_i\frac{p_{_i}^{2-q}+p_{_i}}{2} \ ,
\end{equation}
and by introducing the function $\overline Z_{_{2-q}}$,
through Eq. (\ref{mz}), we obtain the relation
\begin{equation}
\left(\overline Z_{_{2-q}}\right)^{q-1}=1+(q-1)\,S_{_{2-q}}
\ ,
\end{equation}
according to the results reported in \cite{Scarfone4}.\\
From definition (\ref{zq}) we readily obtain the
$q$-deformed thermodynamical potentials corresponding to
the Massieu potential $\Phi_{_{2-q}}$ and to the free
energy $F_{_{2-q}}$, given respectively by
\begin{eqnarray}
&&\Phi_{_{2-q}}=
\ln_{_{2-q}}\Big(Z_{_{2-q}}\Big)\equiv{2\over3-q}\Big({\cal I}_{_{2-q}}+\gamma\Big)+{q-1\over3-q}\,\beta\,U \ ,\\
&&F_{_{2-q}}=-{1\over\beta}\,\ln_{_{2-q}}\Big(Z_{_{2-q}}\Big)\equiv-{2\over(3-q)\,\beta}\Big({\cal
I}_{_{2-q}}+\gamma\Big)-{q-1\over3-q}\,U \ .
\end{eqnarray}
%%%%%%%%%%%%%%%%%%%%%%%%%%%%%%%%%%%%%%%%%%%%%%%%%%%%%%%%%%%%%%%%%%%%%%%%%%%%%%%%%%%%%%
\section{Conclusions}

In the present work we have analyzed some aspects of the
thermostatistics theory based on the two-parameter deformed
Sharma-Taneja-Mittal entropy. In particular, we have
studied the Legendre structure of the theory by introducing
consistently some generalized thermodynamical functions
like the canonical partition function, the free energy and
the Massieu potential, and we have analyzed the
relationships among these functions and the entropy. All
the theoretical structure collapse, in the
$(\kappa,\,r)\to(0,\,0)$ limit, to the standard theory
based on the BG entropy.\\
We recall that the pdf associated to the thermodynamical
potentials introduced in this paper is characterized by an
asymptotic power law behavior. For this reason, the theory
under scrutiny is expected to be relevant in the study of
the thermostatistics proprieties of those complex systems
exhibiting such behavior in the observed
pdf.\\

%%%%%%%%%%%%%%%%%%%%%%%%%%%%%%%%%%%%%%%%%%%%%%%%%%%%%%%%%%%%%%%%%%%%%%%%%%%%%%%%%%%%%%

\noindent {\bf Acknowledgements}\\

\noindent The author wishes to thanks Dr. T. Wada for
useful and stimulating discussions during the preparation
of this manuscript.

%%%%%%%%%%%%%%%%%%%%%%%%%%%%%%%%%%%%%%%%%%%%%%%%%%%%%%%%%%%%%%%%%%%%%%%%%%%%%%%%%%%%%%

\end{document}